# Low Complexity Space-Frequency MIMO OFDM System for Double-Selective Fading Channels

[†]F. Kalbat, *Student Member, IEEE*, and [‡]A. Al-Dweik, *Senior Member, IEEE*

**Abstract**

This paper presents a highly robust space-frequency block coded (SFBC) multiple-input multiple-output (MIMO) orthogonal frequency division multiplexing (OFDM) system. The proposed system is based on applying a short block length Walsh Hadamard transform (WHT) after the SFBC encoder. The main advantage of the proposed system is that the channel frequency responses over every two adjacent subcarriers become equal. Such interesting result provides an exceptional operating conditions for SFBC-OFDM systems transmitting over time and frequency-selective fading channels. Monte Carlo simulation is used to evaluate the bit error rate (BER) performance of the proposed system using various wireless channels with different degrees of frequency selectivity and Doppler spreads. The simulation results demonstrated that the proposed scheme substantially outperforms the standard SFBC-OFDM and the space-time block coded (STBC) OFDM systems in severe time-varying frequency-selective fading channels. Moreover, the proposed system has very low complexity because it is based on short block length WHT.

**Index Terms**

OFDM, Alamouti, space-time, STBC, space-frequency, SFBC, MIMO, precoding.

[†] Department of Electrical and Computer Engineerimg, Khalifa University, UAE, e-mail: fatma.kalbat@kustar.ac.ae.

[‡]Department of Electrical and Computer Engineering, Faculty of Engineering, Western University, London, Ontario, N6A 5B9, Canada, e-mail: aaldweik@uwo.ca, dweik@fulbrightmail.org.

# I. INTRODUCTION

Orthogonal frequency division multiplexing (OFDM) is a multicarrier modulation technique that employs orthogonal subcarriers. OFDM systems can be implemented efficiently by means of inverse fast Fourier transform (IFFT) and fast Fourier transform (FFT) at the transmitter and receiver, respectively. Bandwidth efficiency and immunity to multipath propagation are the main advantages of OFDM over single carrier transmission [1]. Consequently, OFDM has been adopted in many digital communication standards such as digital audio broadcasting (DAB) [2], digital video broadcasting-terrestrial (DVB-T) [3], Interoperability for Microwave Access (WiMAX) technologies [4] and the 4G LTE-Advanced [5].

Despite its immunity against inter-symbol-interference (ISI), OFDM systems do not posses any special immunity against severe frequency-selective channels. If one or more subcarriers are attenuated by a deep fade, such subcarriers can not be recovered which leads to a substantial bit error rate (BER) degradation. Therefore, extensive research have been conducted to overcome this drawback of OFDM systems. In the literature, several solutions are proposed to enhance the performance of OFDM systems in severe frequency-selective channels. However, remarkable attention is given to particular techniques such as precoding, multiple-input multiple-output (MIMO) systems, or hybrid versions that combine precoding and MIMO systems. In general, precoding techniques employ a particular unitary transform such as the Walsh Hadamard transform (WHT) to provide frequency diversity by mixing the frequency domain symbols prior the IFFT [6]-[8]. If a small number of subcarriers is severely attenuated by a deep fade, such subcarriers can still be recovered from other non faded subcarriers. However, the number of lost subcarriers in severe fading channels is large, and hence it will be difficult to recover them. As an alternative solution to precoding, MIMO systems based on Alamouti [9] space-time block codes (STBC) have been considered in the literature [10]-[12]. The main limitation of STBC-OFDM systems is the sensitivity to channel variation in time domain, which causes severe performance degradation due to the channel difference over the STBC block duration. As a remedy for this problem, the STBC symbols can be used to modulate the OFDM subcarriers in frequency domain, thus they are denoted as space-frequency block codes (SFBC) [13]. The SFBC-OFDM may have good BER performance in time-varying channels given that the channel frequency response is not highly frequency-selective. Otherwise, the SFBC will face the same problem of the STBC, but in frequency domain. Therefore, both STBC and SFBC are not expected to provide reliable performance in double-selective fading channels. Some of the main solutions reported in the literature to overcome the double-selective channels challenge are based on hybrid systems that combine precoding and MIMO [12], [14]-[18],.

In addition to the above mentioned systems, adaptive modulation (AM) techniques are considered as an essential tool to optimize the performance of various communication systems in fading channels. AM techniques provide a high degree of freedom for service providers that allow them to maximize the throughput, minimize power consumption and satisfy the quality of service (QoS) requirements [4]. A pivotal motivation for adapting AM techniques for several wireless systems such as WiMax and LTE is to achieve the long promised anytime anywhere (AA) communications. Achieving the AA goal is crucial for both the service providers and end users, even if that is at the expense of the data rate. Consequently, most current wireless standards have included the binary phase shift keying (BPSK) as the last resort to maintain the end user connectivity to the network [4], [5]. Moreover, BPSK is widely used for signaling, synchronization, and channel estimation [3], hence it is quite important to receive such symbols with high reliability.

Based on the discussion in the previous paragraph, it is expected that BPSK combined with MIMO-STBC and OFDM will offer a highly robust system for communication over quasi-static severe frequency-selective channels [19]. However, if the channel changes noticeably over the transmission period of the STBC blocks, and the channel is not severely frequency-selective, SFBC-OFDM will be a better choice [20]. In general, it might be feasible, but quite complex, to switch between the two systems based on the channel status. Moreover, ubiquitous and mobile computing applications usually experience both time and frequency-selective channels, i.e., double-selective channels. In such channels both the STBC and SFBC will suffer from a degraded performance even when BPSK is adopted. Hence, switching between the two configurations may not offer any noticeable performance enhancement.

Since SFBC-OFDM performs well in time-selective channels with low levels of frequencies selectivity, we propose in this paper a new technique to overcome the frequency selectivity limitation of the channel. Consequently, the system will be highly robust in double-selective channels. Unlike other solutions reported in the literature, the proposed system guarantees that the channel frequency responses over two adjacent subcarriers in SFBC-OFDM systems are identical. Consequently, the performance degradation due to the difference of channel coefficients will be completely eliminated. The proposed system is based on a simple $2 \times 2$ WHT, hence it can be considered as a low complexity solution. The optimum performance of the proposed system is achieved with BPSK modulation, which is widely adopted in AM systems to maintain the end user connected in severe transmission conditions such as double-selective channels. Furthermore,

BPSK is widely used for signaling, synchronization and channel estimation, therefore it is highly desired to receive such symbols reliably.

The remainder of the paper is organized as follows. The OFDM-STBC and SFBC system models are described in Section II. The proposed WHT-SFBC system is presented in Section III. Numerical results are presented in Section IV. Finally, the conclusions are given in Section V.

Unless it is otherwise specified, the notations used in this paper are as follows. Uppercase boldface and blackboard letters such as $\mathbf{H}$ and $\mathbb{H}$ will denote $N \times N$ matrices, whereas lowercase boldface letters such as $\mathbf{x}$ will denote row or column vectors with $N$ elements. Bold upper and lowercase letters with a tilde such as $\tilde{\mathbf{H}}$ and $\tilde{\mathbf{r}}$ will denote $N_t \times N_t$ matrices and row/column vectors with $N_t$ elements, respectively. Bold upper case or lowercase letters with a subscript such as $\mathbf{H}_k$ and $\mathbf{r}_k$ will denote $2 \times 2$ and $1 \times 2$ submatrices and subvectors, respectively. The complex conjugate and the transpose of $\mathbf{x}$ are denoted as $\mathring{\mathbf{x}}$ and $\mathbf{x}^T$, respectively.

## II. STBC/SFBC OFDM System Model

### A. STBC-OFDM

In this work we consider an Alamouti coded MIMO OFDM system that consists of two transmit and two receive antennas [9]. If the channel is time invariant over two consecutive OFDM symbols, the STBC encoder output can be transmitted as STBC-OFDM [18]. Therefore, the STBC encoder output for a two consecutive data sequences $\mathbf{a}^1$ and $\mathbf{a}^2$ each of which consists of $N$ data symbols can be expressed as [9],

$$\begin{bmatrix} \mathbf{a}^1 & \mathbf{a}^2 \end{bmatrix} \stackrel{STBC}{\rightarrow} \begin{bmatrix} \mathbf{a}^1 & -\mathring{\mathbf{a}}^2 \\ \mathbf{a}^2 & \mathring{\mathbf{a}}^1 \end{bmatrix} = \begin{bmatrix} \mathbf{u}^{1,1} & \mathbf{u}^{1,2} \\ \mathbf{u}^{2,1} & \mathbf{u}^{2,2} \end{bmatrix}, \tag{1}$$

where $\mathbf{u}^{s,\tau}$ denotes the transmitted block from antenna $s$ at time period $\tau$, $\{s,\tau\} \in \{1,2\}$. The data symbols of sequences $\mathbf{a}^1$ and $\mathbf{a}^2$ are usually selected uniformly from a general constellation such as $M$-ary phase shift keying (MPSK) or quadrature amplitude modulation (QAM). However before transmission, each of the four sequences is OFDM modulated by applying an $N$-points IFFT to each sequence at the output of the STBC encoder. The IFFT output can be written as

$$\mathbf{x}^{s,\tau} = \mathbf{F}^H \mathbf{u}^{s,\tau}, \tag{2}$$

where $\mathbf{F}^H$ is the Hermetian transpose of the normalized $N \times N$ FFT matrix. The elements of $\mathbf{F}^H$ are defined as $F_{i,k}^H = (1/\sqrt{N})e^{j2\pi ik/N}$ where $i$ and $k$ denote the row and column numbers $\{i,k\} = 0, 1, ...,$

$N-1$, respectively. Consequently, the $n$th sample in $\mathbf{x}^{s,\tau}$ can be expressed as,

$$x_n^{s,\tau} = \frac{1}{\sqrt{N}} \sum_{i=0}^{N-1} u_i^{s,\tau} e^{\frac{j2\pi ni}{N}}, \qquad n = 0, 1,..., N-1. \tag{3}$$

To prevent inter-symbol-interference (ISI) and maintain the subcarriers' orthogonality in frequency-selective multipath fading channels, a cyclic prefix (CP) of length $P$ samples is appended as a preamble at the beginning of each IFFT output to form a complete OFDM symbol $\tilde{\mathbf{x}}^{s,\tau}$ that has $N_t = N + P$ samples, and a duration of $T_t$ seconds. The OFDM symbols are upconverted to higher frequencies and transmitted through the corresponding antenna and time slot as described in Table I. Hence, the complex baseband OFDM symbol can be expressed as

$$\tilde{\mathbf{x}}^{s,\tau} = [x_{N-P}^{s,\tau},\, x_{N-P+1}^{s,\tau},..., x_{N-1}^{s,\tau},\, x_0^{s,\tau},\, x_1^{s,\tau},\ldots, x_{N-1}^{s,\tau}]^T. \tag{4}$$

At the receiver front-end, the received signal is down-converted to baseband and sampled at a rate $T_s = T_t/N_t$. In this work we assume that the channels between the transmit antenna $s$ and receive antenna $v$ are independent and uncorrelated Rayleigh fading channels. Each of the channels is composed of $L_h^{s,\tau}+1$ independent multipath components each of which has a gain $h_l^{s,v}$ and delay $l^{s,\tau} \times T_s$, where $l^{s,} \in \{0, 1,..., L_h^{s,\tau}\}$. To deduce the system mathematical model, it is assumed that the channel taps are constant over one OFDM symbol period [21], [22], and it is assumed that $L_h^{s,\tau} \leq P$ as well. Therefore, the received sequence at receiver $v$ during time period $\tau$ can be expressed as

$$\tilde{\mathbf{y}}^{v,\tau} = \tilde{\mathbb{H}}^{1,v,\tau}\tilde{\mathbf{x}}^{1,\tau} + \tilde{\mathbb{H}}^{2,v,\tau}\tilde{\mathbf{x}}^{2,\tau} + \tilde{\mathbf{z}}^{v,\tau}, \tag{5}$$

where $\tilde{\mathbb{H}}^{s,v,\tau}$ is the channel matrix between transmit antenna $s$ and receive antenna $v$ at time slot $\tau$, is an $N_t \times N_t$ Toeplitz matrix with $h_0^{s,\tau}$ on the principal diagonal and $h_1^{s,\tau},..., h_{L_h}^{s,\tau}$ on the minor diagonals, respectively. The noise vector $\tilde{\mathbf{z}}$ is modeled as a complex white Gaussian noise process with zero mean and variance $\sigma_z^2 = E[|z_n|^2]$. Subsequently, the receiver should identify and discard the $P$ CP samples, which yields the sequences $\mathbf{y}^{v,\tau} = [y_0^{s,\tau},\, y_1^{s,\tau},\, \ldots,\, y_{N-1}^{s,\tau}]$,

$$\mathbf{y}^{v,\tau} = \mathbb{H}^{1,v,\tau}\mathbf{x}^{1,\tau} + \mathbb{H}^{2,v,\tau}\mathbf{x}^{2,\tau} + \mathbf{z}^{v,\tau}. \tag{6}$$

The channel matrix $\mathbb{H}$ is an $N \times N$ circulant matrix. By noting that $\mathbf{F}\,\mathbb{H}^{s,v,\tau}\mathbf{x}^{s,\tau} = \mathbf{F}\,\mathbb{H}^{s,v,\tau}\mathbf{F}^H\mathbf{u}^{s,\tau} = \mathbf{H}^{s,v,\tau}\mathbf{u}^{s,\tau}$,

the FFT output can be expressed as

$$\mathbf{r}^{v,\tau} = \mathbf{H}^{1,v,\tau}\mathbf{u}^{1,\tau} + \mathbf{H}^{2,v,\tau}\mathbf{u}^{2,\tau} + \boldsymbol{\eta}^{v,\tau}, \tag{7}$$

where $\boldsymbol{\eta} = \mathbf{F}\mathbf{z}$ denotes the noise vector that has the same properties of the original vector $\mathbf{z}$, and $\mathbf{H}$ is a diagonal matrix that represents the channel frequency response,

$$\mathbf{H} = \mathrm{diag}\left([H_0, H_1, \cdots, H_{N-1}]\right), \tag{8}$$

where $H_k = \sum_{m=0}^{L_h} h_m e^{-j2\pi mk/N}$. Assuming that the channel remains constant over two consecutive OFDM symbols, the time slot index in (7) can be dropped from the channel matrix, hence (7) can be written as

$$\mathbf{r}^{v,\tau} = \mathbf{H}^{1,v}\mathbf{u}^{1,\tau} + \mathbf{H}^{2,v}\mathbf{u}^{2,\tau} + \boldsymbol{\eta}^{v,\tau}, \tag{9}$$

which can be expanded to

$$\mathbf{r}^{1,1} = \mathbf{H}^{1,1}\mathbf{a}_1 + \mathbf{H}^{2,1}\mathbf{a}^2 + \boldsymbol{\eta}^{1,1}$$

$$\mathbf{r}^{2,1} = \mathbf{H}^{1,2}\mathbf{a}_1 + \mathbf{H}^{2,2}\mathbf{a}^2 + \boldsymbol{\eta}^{2,1}$$

$$\mathbf{r}^{1,2} = -\mathbf{H}^{1,1}\mathring{\mathbf{a}}^2 + \mathbf{H}^{2,1}\mathring{\mathbf{a}}^1 + \boldsymbol{\eta}^{1,2}$$

$$\mathbf{r}^{2,2} = -\mathbf{H}^{1,2}\mathring{\mathbf{a}}^2 + \mathbf{H}^{2,2}\mathring{\mathbf{a}}^1 + \boldsymbol{\eta}^{2,2} \tag{10}$$

Therefore, we can write the $k$th subcarrier of $\mathbf{r}^{v,\tau}$

$$\begin{bmatrix} r_k^{1,1} \\ r_k^{2,1} \\ \mathring{r}_k^{1,2} \\ \mathring{r}_k^{2,2} \end{bmatrix} = \begin{bmatrix} H_k^{1,1} & H_k^{2,1} \\ H_k^{1,2} & H_k^{2,2} \\ -\mathring{H}_k^{1,1} & -\mathring{H}_k^{2,1} \\ -\mathring{H}_k^{1,2} & -\mathring{H}_k^{2,2} \end{bmatrix} \begin{bmatrix} a_k^1 \\ a_k^2 \end{bmatrix} + \begin{bmatrix} \eta_k^{1,1} \\ \eta_k^{2,1} \\ \mathring{\eta}_k^{1,2} \\ \mathring{\eta}_k^{2,2} \end{bmatrix}$$

$$\acute{\mathbf{r}}_k = \acute{\mathbf{H}}_k \mathbf{a}_k + \acute{\boldsymbol{\eta}}_k. \tag{11}$$

Finally, the standard STBC decoder output $\acute{\mathbf{a}}_k$ can be expressed as

$$\acute{\mathbf{a}}_k = \left[\acute{\mathbf{H}}_k^H \acute{\mathbf{H}}_k\right]^{-1} \acute{\mathbf{H}}_k^H \acute{\mathbf{r}}_k. \tag{12}$$

## B. SFBC-OFDM

In the case that the channel varies significantly over two consecutive OFDM symbols, then SFBC-OFDM can be utilized as an efficient replacement for the STBC [23]. In SFBC-OFDM, a single data sequence $\mathbf{a} = [a_0, a_1, ..., a_{N-1}]$ is applied to a standard Alamouti STBC encoder where the output is mapped as follows,

$$\begin{bmatrix} a_k & a_{k+1} \end{bmatrix} \xrightarrow{STBC} \begin{bmatrix} a_k & -\mathring{a}_{k+1} \\ a_{k+1} & \mathring{a}_k \end{bmatrix}, \quad k = 0, 2, ..., N-2. \tag{13}$$

Then each row in (13) is buffered to form an $N$ symbols sequence,

$$\mathbf{b}^1 = [a_0, \ -\mathring{a}_1, \ a_2, \ -\mathring{a}_3, ..., \ a_{N-2}, \ -\mathring{a}_{N-1}]$$
$$\mathbf{b}^2 = [a_1, \ \mathring{a}_0, \ a_3, \ \mathring{a}_2, ..., \ a_{N-1}, \ \mathring{a}_{N-2}]. \tag{14}$$

Similar to the STBC-OFDM, the sequences $\mathbf{b}^1$ and $\mathbf{b}^2$ are applied to an OFDM modulator and upconverter, then $\mathbf{b}^1$ is transmitted through antenna 1 and $\mathbf{b}^2$ is transmitted through antenna 2. Following the same procedure for the STBC-OFDM case, the received signals after the FFT process at receiver $v$ can be expressed

$$\mathbf{r}^v = \mathbf{H}^{1,v}\mathbf{b}^1 + \mathbf{H}^{2,v}\mathbf{b}^2 + \boldsymbol{\eta}^v. \tag{15}$$

Since the matrices $\mathbf{H}^{s,v}$ are diagonal, thus each two consecutive elements in $\mathbf{r}^v$ will have the same form as (15). Consequently, by substituting $b_k^1 = a_k$, $b_k^2 = a_{k+1}$, $b_{k+1}^1 = -\mathring{a}_{k+1}$, $b_{k+1}^2 = \mathring{a}_k$, the FFT output at any two consecutive subcarriers $k$ and $k+1$ can be expressed as

$$r_k^v = H_k^{1,v} a_k + H_k^{2,v} a_{k+1} + \eta_k^v$$

$$r_{k+1}^v = -H_{k+1}^{1,v} \mathring{a}_{k+1} + H_{k+1}^{2,v} \mathring{a}_k + \eta_{k+1}^v.$$

Assuming that the channel remains constant over two adjacent subcarriers, i.e. $H_k^{1,v} = H_{k+1}^{1,v}$, the FFT output can be expressed as

$$\begin{bmatrix} r_k^1 \\ r_k^2 \\ \mathring{r}_{k+1}^1 \\ \mathring{r}_{k+1}^2 \end{bmatrix} = \mathbf{\acute{H}}_k \begin{bmatrix} a_k \\ a_{k+1} \end{bmatrix} + \begin{bmatrix} \eta_k^1 \\ \eta_k^2 \\ \mathring{\eta}_{k+1}^1 \\ \mathring{\eta}_{k+1}^2 \end{bmatrix}, \tag{16}$$

which can be written in vector notations as

$$\acute{\mathbf{r}}_k = \acute{\mathbf{H}}_k \, \acute{\mathbf{a}}_k + \acute{\boldsymbol{\eta}}_k, \tag{17}$$

where $\acute{\mathbf{H}}_k$ is similar to the channel matrix of the STBC-OFDM given in (11). Therefore, the information symbols can be simply extracted using the standard STBC decoder [9],

$$\acute{\mathbf{a}}_k = \left[ \acute{\mathbf{H}}_k^H \, \acute{\mathbf{H}}_k \right]^{-1} \acute{\mathbf{H}}_k^H \, \acute{\mathbf{r}}_k. \tag{18}$$

## III. THE PROPOSED SYSTEM

As it can be noted from the previous subsection, the SFBC has also the requirement for the channel equality, but it is in frequency domain. In order to use the standard Alamouti decoder with minimum performance degradation, the channel frequency response over each pair of adjacent subcarriers modulated by the symbols $\{a_k, -\mathring{a}_{k+1}\}$ and $\{a_{k+1}, \mathring{a}_k\}$ at transmitters 1 and 2, respectively, should be identical. Otherwise, the system performance may suffer a serious BER degradation. Although the equal channel frequency response condition might be satisfied in flat and slow fading channels, the case is drastically different in severe frequency-selective channels. Consequently, similar to the STBC-OFDM system, the performance of the SFBC-OFDM is highly dependent on the channel conditions.

In [6], McCloud proved that if each group of subcarriers in an OFDM system are precoded with a short length unitary transform whose elements have equal magnitudes, then each subcarrier will experience the same instantaneous signal-to-noise ratio (SNR). In this paper we introduce an important extension of McCloud's work when BPSK modulation is adopted. As it is shown later in the paper, using a $2 \times 2$ unitary transform as a precoder in an OFDM system whose subcarriers are modulated by real data symbols with equal absolute values results in an identical channel frequency response over each pair of subcarriers involved in precoding. Although this fact may not be of great value for SISO systems, it is actually priceless for MIMO SFBC-OFDM systems. In fact this result guarantees that the channel equality condition is satisfied regardless of the frequency selectivity level of the channel. As a result, the proposed system will be highly robust over frequency-selective channels. Moreover, since the SFBC is robust against time-varying channels, the proposed system will be robust in double-selective channels.

The proposed SFBC-WHT is constructed by first dividing each of the sequences $\mathbf{b}^1$ and $\mathbf{b}^2$ given in

(14) into $N/2$ blocks of symbols, where each block consists of two symbols,

$$\mathbf{c}_0^1 = \begin{bmatrix} a_0 \\ -\mathring{a}_1 \end{bmatrix}, \mathbf{c}_1^1 = \begin{bmatrix} a_2 \\ -\mathring{a}_3 \end{bmatrix}, ..., \mathbf{c}_{\frac{N}{2}-1}^1 = \begin{bmatrix} a_{N-2} \\ -\mathring{a}_{N-1} \end{bmatrix},$$

$$\mathbf{c}_0^2 = \begin{bmatrix} a_1 \\ \mathring{a}_0 \end{bmatrix}, \mathbf{c}_1^2 = \begin{bmatrix} a_3 \\ \mathring{a}_2 \end{bmatrix}, ..., \mathbf{c}_{\frac{N}{2}-1}^2 = \begin{bmatrix} a_{N-1} \\ \mathring{a}_{N-2} \end{bmatrix}.$$

Hence, we can write

$$\mathbf{c}_k^1 = \begin{bmatrix} a_k \\ -\mathring{a}_{k+1} \end{bmatrix}, \mathbf{c}_k^2 = \begin{bmatrix} a_{k+1} \\ \mathring{a}_k \end{bmatrix}, \quad k = 0, 2, ..., N-2.$$

Each of the blocks $\mathbf{c}_k^1$ and $\mathbf{c}_k^2$ is applied to a $2 \times 2$ WHT defined as

$$\mathbf{T}_2 = \frac{1}{\sqrt{2}} \begin{bmatrix} 1 & 1 \\ 1 & -1 \end{bmatrix},$$

which gives

$$\mathbf{d}_k^1 = \mathbf{T}_2 \mathbf{c}_k^1 = \frac{1}{\sqrt{2}} \begin{bmatrix} a_k - \mathring{a}_{k+1} \\ a_k + \mathring{a}_{k+1} \end{bmatrix},$$

$$\mathbf{d}_k^2 = \mathbf{T}_2 \mathbf{c}_k^2 = \frac{1}{\sqrt{2}} \begin{bmatrix} a_{k+1} + \mathring{a}_k \\ a_{k+1} - \mathring{a}_k \end{bmatrix}. \tag{19}$$

The $N/2$ blocks are then concatenated to form the vectors $\mathbf{d}^1$ and $\mathbf{d}^2$, where $\mathbf{d}^k = \begin{bmatrix} \mathbf{d}_0^k & \mathbf{d}_1^k & \cdots & \mathbf{d}_{\frac{N}{2}-1}^k \end{bmatrix}^T$.

The remaining parts of the transmission and the reception processes are identical to the regular SFBC-OFDM described in subsection II.B except that $\mathbf{b}^s$ is replaced by $\mathbf{d}^s$, $s \in \{1, 2\}$. Thus, the received signal at receiver $v$ can be expressed as

$$\mathbf{r}^v = \mathbf{H}^{1,v} \mathbf{d}^1 + \mathbf{H}^{2,v} \mathbf{d}^2 + \boldsymbol{\eta}^v. \tag{20}$$

However, because the channel matrices $\mathbf{H}^{s,v}$ are diagonal, each two adjacent subcarriers can be extracted before the SFBC decoding process. Thus, the $k$th block that consists of the samples $r_k^v$ and $r_{k+1}^v$ can be expressed as,

$$\mathbf{r}_k^v = \begin{bmatrix} H_k^{1,v} & 0 \\ 0 & H_{k+1}^{1,v} \end{bmatrix} \begin{bmatrix} d_k^1 \\ d_{k+1}^1 \end{bmatrix} + \begin{bmatrix} H_k^{2,v} & 0 \\ 0 & H_{k+1}^{2,v} \end{bmatrix} \begin{bmatrix} d_k^2 \\ d_{k+1}^2 \end{bmatrix} + \begin{bmatrix} \eta_k^v \\ \eta_{k+1}^v \end{bmatrix}, \quad k = 0, 2, ..., N-2, \quad (21)$$

Since the inverse WHT and the WHT are identical, i.e., $\mathbf{T}_2^H = \mathbf{T}_2$, we can apply the WHT to the FFT output,

$$\mathbf{w}_k^v = \mathbf{T}_2 \mathbf{r}_k^v \quad (22)$$

$$= \frac{1}{\sqrt{2}} \begin{bmatrix} H_k^{1,v} & H_{k+1}^{1,v} \\ H_{k+1}^{1,v} & -H_{k+1}^{1,v} \end{bmatrix} \begin{bmatrix} d_k^1 \\ d_{k+1}^1 \end{bmatrix} + \frac{1}{\sqrt{2}} \begin{bmatrix} H_k^{2,v} & H_{k+1}^{2,v} \\ H_k^{2,v} & -H_{k+1}^{2,v} \end{bmatrix} \begin{bmatrix} d_k^2 \\ d_{k+1}^2 \end{bmatrix} + \begin{bmatrix} \acute{\eta}_k^v \\ \acute{\eta}_{k+1}^v \end{bmatrix}, \quad (23)$$

where $\acute{\boldsymbol{\eta}}^v = \mathbf{T}_2 \boldsymbol{\eta}^v$. However, by noting that the WHT output $d_k^s$ and $d_{k+1}^s$ for the special case of BPSK $a_k \in \{1, -1\}$, which are given in Table II, we notice that for every two adjacent subcarriers, one of them should be zero, i.e.,

$$\text{if } d_k^s \neq 0 \rightarrow d_{k+1}^s = 0.$$

$$\text{if } d_{k+1}^s \neq 0 \rightarrow d_k^s = 0. \quad (24)$$

To evaluate the effect of this unique property, consider first the case where $a_k = a_{k+1}$. For this case we have

$$\mathbf{c}_k^1 = \begin{bmatrix} a_k \\ -a_k \end{bmatrix}, \mathbf{d}_k^1 = \mathbf{T}_2 \mathbf{c}_k^1 = \frac{1}{\sqrt{2}} \begin{bmatrix} a_k - a_{k+1} \\ a_k + a_{k+1} \end{bmatrix} = \frac{1}{\sqrt{2}} \begin{bmatrix} 0 \\ a_k \end{bmatrix}, \quad (25)$$

$$\mathbf{c}_k^2 = \begin{bmatrix} a_k \\ a_k \end{bmatrix}, \mathbf{d}_k^2 = \mathbf{T}_2 \mathbf{c}_k^2 = \frac{1}{\sqrt{2}} \begin{bmatrix} a_k + a_{k+1} \\ a_{k+1} - a_k \end{bmatrix} = \frac{1}{\sqrt{2}} \begin{bmatrix} a_k \\ 0 \end{bmatrix}. \quad (26)$$

Moreover, note that $d_k^1 = d_{k+1}^2 = 0$. Therefore we can write (22) as

$$\mathbf{w}_k^v = \begin{bmatrix} H_k^{1,v} & H_{k+1}^{1,v} \\ H_k^{1,v} & -H_{k+1}^{1,v} \end{bmatrix} \begin{bmatrix} 0 \\ a_k \end{bmatrix} + \begin{bmatrix} H_k^{2,v} & H_{k+1}^{2,v} \\ H_k^{2,v} & -H_{k+1}^{2,v} \end{bmatrix} \begin{bmatrix} a_k \\ 0 \end{bmatrix} + \begin{bmatrix} \grave{\eta}_k^v \\ \grave{\eta}_{k+1}^v \end{bmatrix}$$

$$= \begin{bmatrix} H_{k+1}^{1,v} a_k \\ -H_{k+1}^{1,v} a_k \end{bmatrix} + \begin{bmatrix} H_k^{2,v} a_k \\ H_k^{2,v} a_k \end{bmatrix} + \begin{bmatrix} \grave{\eta}_k^v \\ \grave{\eta}_{k+1}^v \end{bmatrix}. \quad (27)$$

However (27) can be rewritten as

$$\mathbf{w}_k^v = \begin{bmatrix} H_{k+1}^{1,v} & 0 \\ 0 & H_{k+1}^{1,v} \end{bmatrix} \begin{bmatrix} a_k \\ -a_k \end{bmatrix} + \begin{bmatrix} H_k^{2,v} & 0 \\ 0 & H_k^{2,v} \end{bmatrix} \begin{bmatrix} a_k \\ a_k \end{bmatrix} + \begin{bmatrix} \acute{\eta}_k^v \\ \acute{\eta}_{k+1}^v \end{bmatrix}$$
$$= \mathbf{\Lambda}_k^{1,v} \mathbf{c}_k^1 + \mathbf{\Lambda}_k^{2,v} \mathbf{c}_k^2 + \acute{\boldsymbol{\eta}}_k^v. \quad (28)$$

It is quite interesting to note that the channel matrices $\mathbf{\Lambda}_k^{1,1}$ and $\mathbf{\Lambda}_k^{2,1}$ are diagonal with identical elements. Therefore, the channel frequency responses at every two adjacent subcarriers are identical. It is straightforward to extend the results of (28) to other case where $a_k = -a_{k+1}$. The channel frequency response for a conventional SFBC-OFDM and the proposed WHT SFBC-OFDM systems is depicted in Fig. 1. The OFDM systems is composed of 64 subcarriers. The channel considered in this example is a static slow fading channel, which is denoted as Ch-1 in Section IV.

## IV. NUMERICAL RESULTS

Monte Carlo simulations are used to evaluate the performance of the proposed MIMO SFBC-WHT system over static and time-varying frequency-selective multipath fading channels. The performance of the proposed system is evaluated in terms of the BER, and it is compared to the performance of the standard STBC and SFBC systems.

The OFDM system considered in this paper has $N = 64$ subcarriers and the number of cyclic prefix samples $P = 16$. Unless it is otherwise specified, all data symbols are BPSK modulated with symbol rate of $4.17$ kbps, and the carrier frequency is equal to $2.4$ GHz. Three multipath fading channel models are considered in this work [21], Ch-1 corresponds to a slow fading channel with a normalized delay vector of $[0, 1, 2, 3, 4]$ samples, average gains of $[0.35, 0.25, 0.18, 0.13, 0.09]$, and mean square delay spread $\sigma^2 = 1.74$. Ch-2 corresponds to moderate frequency-selective channel having 5 taps with normalized delays of $[0, 1, 2, 6, 11]$ samples and average gains $[0.34, 0.28, 0.23, 0.11, 0.04]$, Ch-3 corresponds to a severe frequency-selective channel having 4 taps with normalized delays of $[0, 4, 8, 12]$ samples and average gains $[0.25, 0.25, 0.25, 0.25]$. The mean-squared delay spread of Ch-2 and Ch-3 is equal to $6.37$ and $20$, respectively. The path gains are generated as complex independent random variables and the fading is generated using the Jakes' Doppler spectrum model. The effect of the channel variation is evaluated using four different Doppler spread $f_d$ values, namely $0$, $42$, $105$ and $210$ Hz, which corresponds to $0$, $1\%$, $2.5\%$ and $5\%$ of the subcarrier frequency spacing. The considered Doppler spreads corresponds to a vehicle speeds of $0$, $19$, $47$ and $95$ km/hour, respectively.

The BER performance of the proposed SFBC-WHT and the standard SFBC systems over channels 1, 2 and 3 is given in Fig. 2 for $f_d = 0$. As it can be noted from this figure, the SFBC-WHT BER is almost independent of the channel frequency selectivity. In contrary to the proposed system, the standard SFBC is highly sensitive to the channel variation in frequency domain, which is demonstrated as BER error floors at moderate and high SNR.

Fig. 3 considers the effect of the Doppler spread combined with a slow fading channel, i.e., Ch-1. It can be noted from this figure that the standard STBC and the proposed system have identical performance for $f_d = 0$. The SFBC BER is slightly degraded since the channel changes slowly in frequency domain. Increasing $f_d$ to 42 Hz increases the BER of all systems due to the inter-carrier-interference (ICI) [24]. However, the proposed system still outperforms the standard SFBC and STBC. At $f_d = 105$ Hz, it can be noted that STBC performance degrades severely due to the channel variation over the STBC block period. The SFBC and the proposed system have experienced some performance degradation as well, however the SFBC performance is worse because it is affected by the channel variation in frequency domain and the ICI due to the channel variation in time domain. The same conclusions can be made for the $f_d = 210$ Hz case, however the SFBC and the proposed system have equivalent performance because the performance is dominated by the ICI [24].

Fig. 4 presents the BER performance of the proposed and other considered systems over the severe fading channel (Ch-3) with low Doppler spreads. As it can be noted from Fig. 4, the SFBC suffers from high error floor due to the channel frequency selectivity, which dominates the performance at low Doppler spread values. The STBC and the proposed system demonstrated better performance however the proposed system outperforms the STBC, for which the BER is increased by the channel variations in time domain.

The BER for moderate and high mobility conditions with the severe fading channel is depicted in Fig. 5. As it can be noted from this figure, the BER of the proposed system for $f_d = 105$ Hz is much lower than that of the SFBC and STBC. On the other hand, the STBC outperforms the SFBC as the channel changes in frequency domain faster than its change in time domain. For $f_d = 210$ Hz, we note the opposite, i.e., the SFBC BER becomes lower than the STBC because the channel changes in time domain faster than in frequency domain. Although all systems suffer from a relatively high error floor values, the proposed system remains the one with the lowest BER. Consequently, the BER degradation of the proposed system results from the ICI caused by the large Doppler spreads rather than the channel variation [24].

Although the proposed system is optimized for BPSK, it actually works for the quadrature phase shift keying (QPSK) modulations as well. However, the channel response equality is not guaranteed for all adjacent subcarriers as in the case of BPSK. For example, if $a_k = \mathring{a}_{k+1}$ or if $a_k = -\mathring{a}_{k+1}$, then $\mathbf{d}^1$ and $\mathbf{d}^2$ will have one of their outputs as a null, which provides identical channel responses over the two subcarriers. The BER performance of the proposed system using QPSK modulation is presented in Fig. 6 for the three considered channels with no Doppler spread. It can be noted from this figure that a non negligible improvement is gained over the standard SFBC, however it is not as large as the one obtained for the BPSK case.

## V. Conclusion

This work presented a novel technique to overcome the channel variation in MIMO SFBC systems. The proposed system enforces the channel responses over each pair of adjacent subcarriers to be identical. Consequently, the performance degradation due to the channel variation is minimized. The proposed system has very low computational complexity due to the use of short block length WHT. Extensive simulation results confirmed the robustness of the proposed system over various channel models with different mobility levels. The proposed system is optimized for BPSK modulation which makes it very attractive option for adaptive modulation systems to guarantee the end user connectivity in severe communication channels.

TABLE I

ANTENNA AND TIME SLOT MAPPING FOR THE STANDARD STBC SYSTEM.

|  | $\tau = 1$ | $\tau = 2$ |
|---|---|---|
| Antenna 1 | $\mathbf{u}^{1,1} = \mathbf{a}^1$ | $\mathbf{u}^{1,2} = -\mathring{\mathbf{a}}^2$ |
| Antenna 2 | $\mathbf{u}^{2,1} = \mathbf{a}^2$ | $\mathbf{u}^{2,2} = \mathring{\mathbf{a}}^1$ |

TABLE II

ANTENNA AND SUBCARRIER MAPPING OF THE PROPSED SYSTEM.

|  |  | $d_k^1$ | $d_{k+1}^1$ | $d_k^2$ | $d_{k+1}^2$ |
|---|---|---|---|---|---|
| $a_k$ | $a_{k+1}$ | $a_k - a_{k+1}$ | $a_k + a_{k+1}$ | $a_{k+1} + a_k$ | $a_{k+1} - a_k$ |
| 1 | 1 | 0 | $\frac{1}{\sqrt{2}}$ | $\frac{1}{\sqrt{2}}$ | 0 |
| -1 | 1 | $-\frac{1}{\sqrt{2}}$ | 0 | 0 | $\frac{1}{\sqrt{2}}$ |
| 1 | -1 | $\frac{1}{\sqrt{2}}$ | 0 | 0 | $-\frac{1}{\sqrt{2}}$ |
| -1 | -1 | 0 | $-\frac{1}{\sqrt{2}}$ | $-\frac{1}{\sqrt{2}}$ | 0 |

.

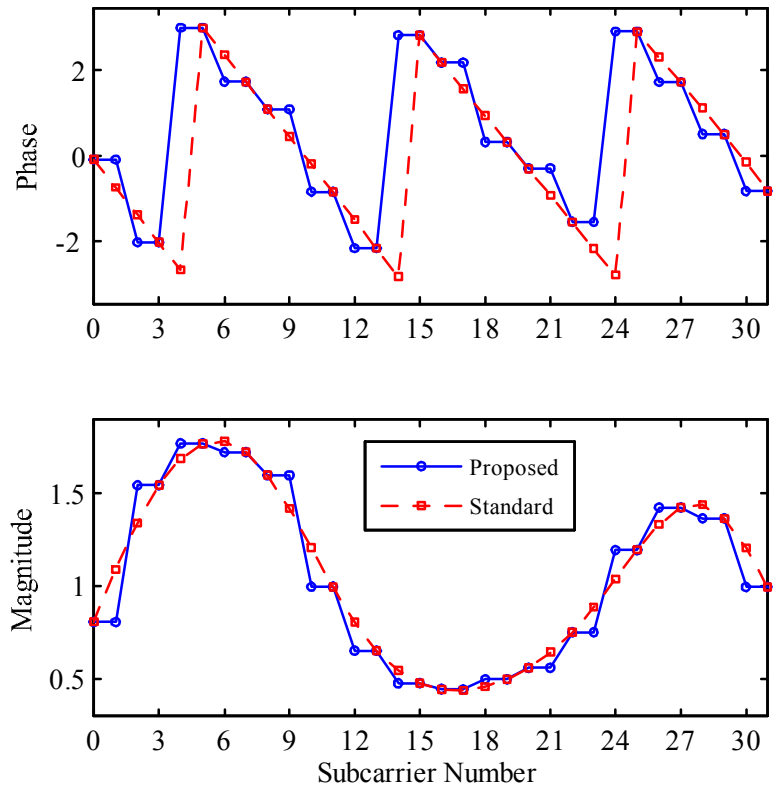

Fig. 1. The phase and magnitude of a sample channel response for the proposed and conventional SFBC-OFDM systems using Ch-1, $f_d = 0$ Hz.

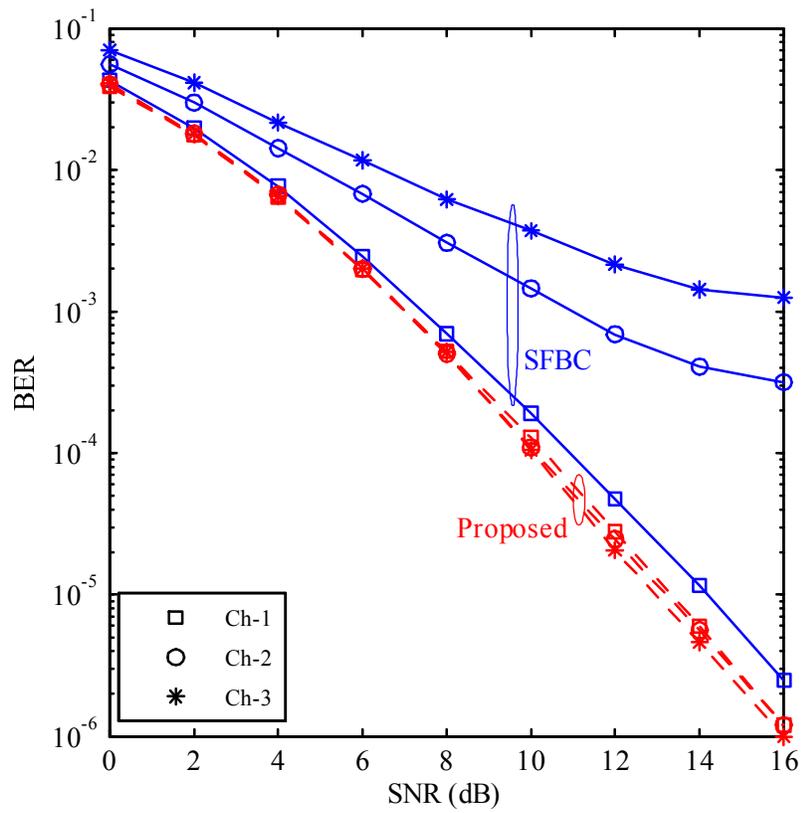

Fig. 2. The BER of the proposed and the standard SFB over various static frequency-selective channel models.

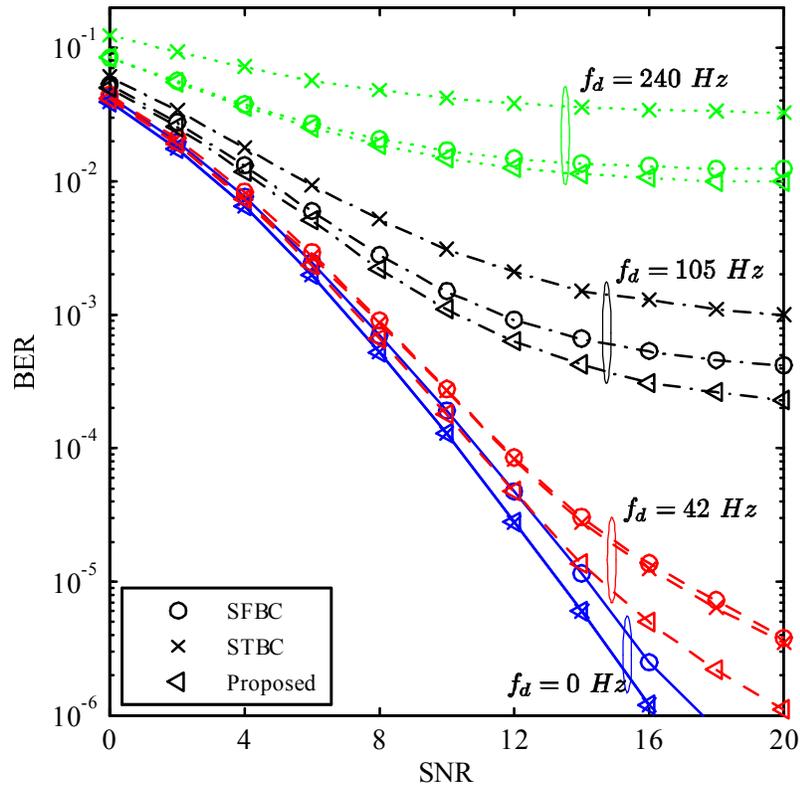

Fig. 3. The BER performance over a slow fading channel (Ch-1) using different Doppler spreads.

20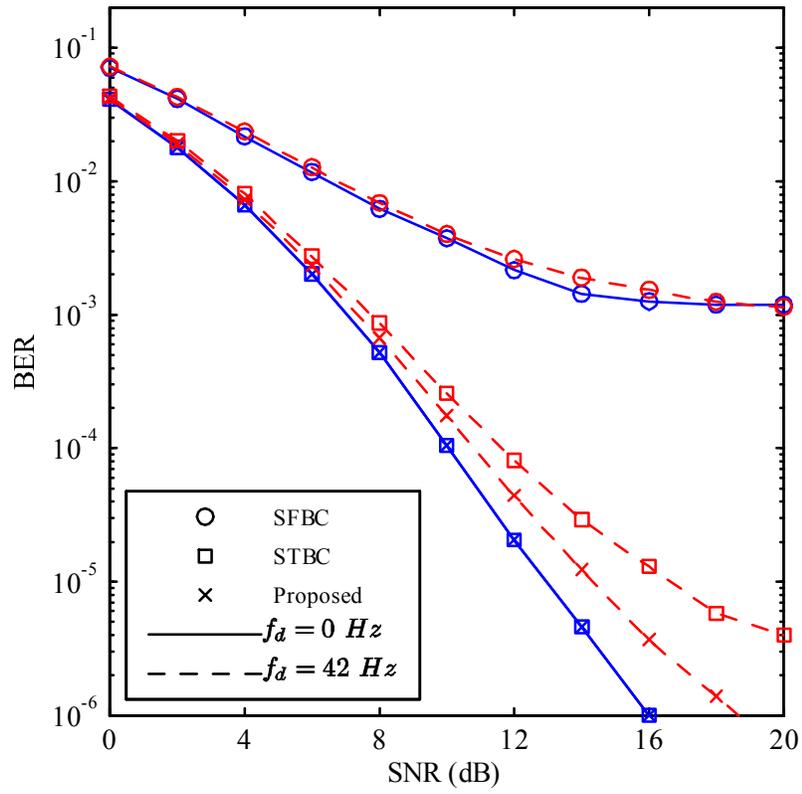

Fig. 4. The BER performance over severe frequency-selective channel (Ch-3) using low Doppler spread values.

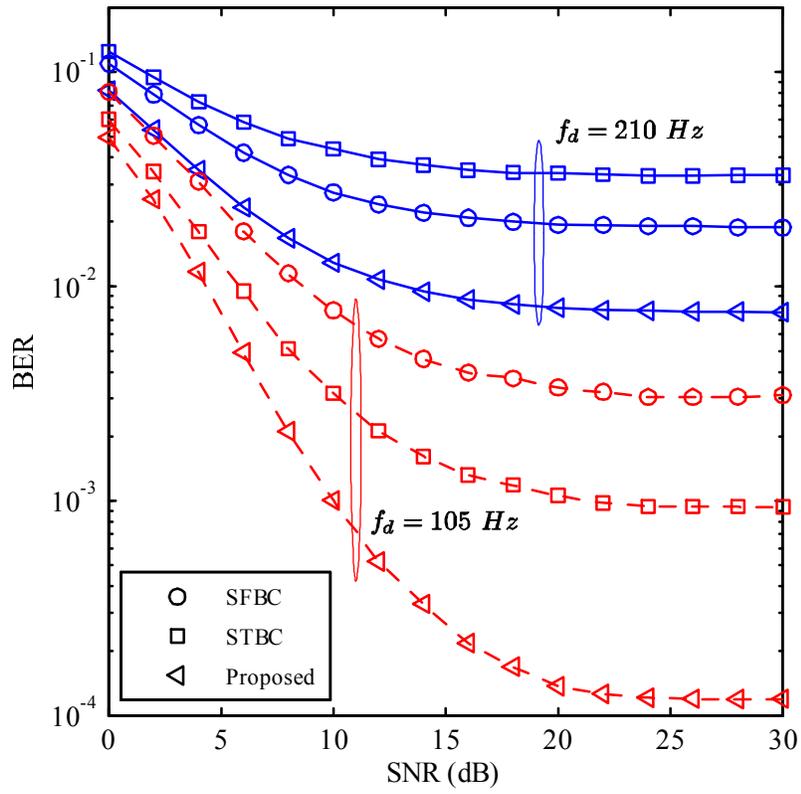

Fig. 5. The BER performance over severe frequency-selective channel (Ch-3) using high Doppler spread values.

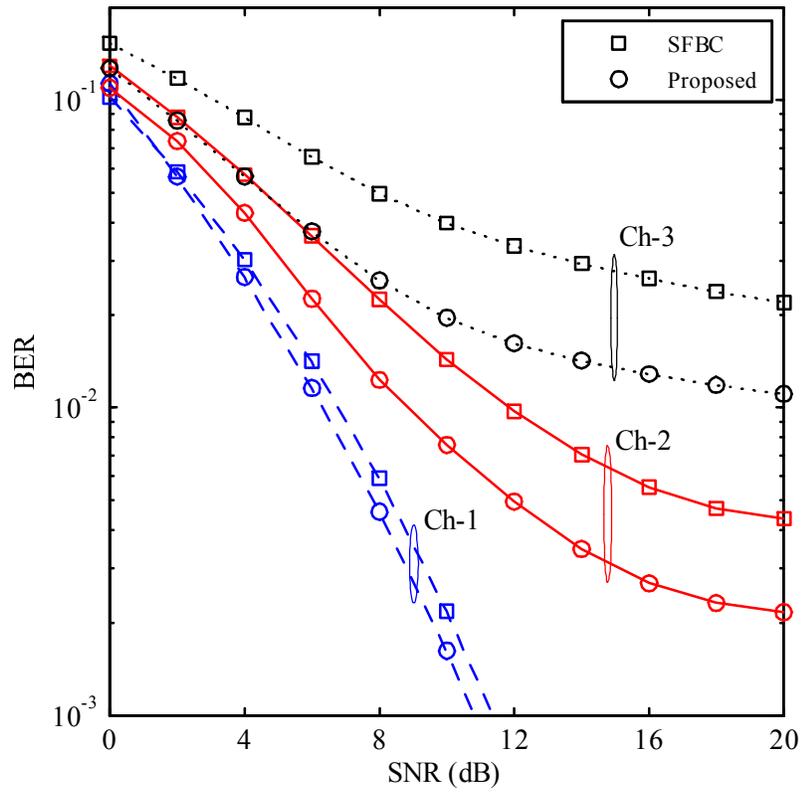

Fig. 6. The BER of the proposed and the standard SFBC over various static frequency-selective channel models using QPSK modulation.